\documentclass[aps,prd,fleqn,superscriptaddress]{revtex4}
\pdfoutput=1
\usepackage{graphicx,color,natbib}
\usepackage{amsmath,amssymb,amsfonts}
\newcommand{\bse}{\begin{subequations}}
\newcommand{\ese}{\end{subequations}}
\newcommand{\be}{\begin{equation}}
\newcommand{\ee}{\end{equation}}
\newcommand{\bea}{\begin{eqnarray}}
\newcommand{\eea}{\end{eqnarray}}
\newcommand{\ba}{\begin{array}}
\newcommand{\ea}{\end{array}}

\input amssym.def
\input amssym.tex

\usepackage[colorlinks=true, linkcolor=blue, bookmarks=true]{hyperref}
\begin{document}
\title{Meson Excitation at Finite Chemical Potential}
\author{A. Hajilou}
\email{{\rm{a}}$_{}$ hajilou@sbu.ac.ir}
\affiliation{Department of Physics, Shahid Beheshti University G.C., Evin, Tehran 19839, Iran}
\author{M. Ali-Akbari}
\email{{\rm{m}}$_{}$ aliakbari@sbu.ac.ir}
\affiliation{Department of Physics, Shahid Beheshti University G.C., Evin, Tehran 19839, Iran}
\begin{abstract}
We consider a probe stable meson in the holographic quark-gluon plasma at zero temperature and chemical potential. Due to the energy injection into the plasma, the temperature and chemical potential are increased to arbitrary finite values and the meson is also excited. Excitation time $t_{ex}$ is the time at which the meson falls into the final excited state. We study the effect of various parameters of theory on the excitation time and observe that for larger values of final temperature and chemical potential the excitation time increases. Furthermore, our outcomes show that the more stable mesons are excited sooner.
\end{abstract}
\maketitle
\tableofcontents
\section{Introduction and results}
Calculating various physical quantities and describing how a time-dependent process evolves in a strongly coupled gauge theory, on the one hand, are always attractive and of course important. On the other hand, standard methods applied in physics problems are mostly based on the perturbative expansion and therefore are not applicable in the model we are interested in due to non-perturbative nature of the theory, i.e. large coupling constant. Quantum chromodynamics is one of the most significant theories we eagerly like to understand its properties at low energies and unfortunately, since its coupling constant is large, there is no unique framework to address our questions. Furthermore, numerical solutions indicate that quark-gluon plasma (QGP) produced at RHIC and LHC is also a strongly coupled plasma and needs a new method to describe its properties \cite{solana}. In short, a suitable candidate is essential to answer various questions of strongly coupled gauge theories.

A remarkable candidate recieved a lot of interest during last decade is AdS/CFT correspondence or more generally gauge-gravity duality \cite{solana, Maldacena}. This duality is a conjectured relation between two physical theories. One of them is a strongly coupled gauge theory in $d$ dimensional space-time and the other one is a classical gravity theory living in an extra dimension of space-time. In fact, parameters, fields and different processes in the gauge theory are translated into appropriate equivalent on the gravity side. For instance, the (thermal) vacuum state on the gauge theory side corresponds to the (black hole-AdS) pure AdS in the gravity theory. Thermalization process, which generally means evolution of a state from zero temperature to a thermal state, is dual to black hole formation in the gravity theory. Moreover, as another example, the meson, quark-antiquark bound state, living in the QGP can be identified with a classical string in the gravity and by using the expectation value of the Wilson loop the static potential between a quark and antiquark has been firstly found in \cite{maldacena1}. For more information, the interested reader is refered to \cite{solana} and references therein.

In this paper, we consider a probe stable meson in the QGP at zero temperature and chemical potential. Then the temperature and chemical potential are simultaneously raised from zero to finite values $T_f$ and $\mu_f$, respectively. Now the questions we would like to answer are how the stable meson reacts to the energy injection into the system and what the characteristics of the new meson state are? Furthermore, it is instructive to know how much time is needed for the meson to fall into the final (excited) state, excitation time, and what the effect of the final values of the temperature and chemical potential is on the excitation time? The holographic dual of the above system is described by the dynamics of a classical string, with appropriate initial and boundary conditions, in the Reissner-Nordstr\"{o}m-AdS Vaidya (RN-AdS-Vaidya) background, as we will review in the next section.
Our main outcomes can be summarized as follows:
\begin{itemize}
\item Making larger each parameter under study in this paper, i.e. final chemical potential, final temperature and the quark-antiquark distance, the bound state is less stable and the energy of the bound state, based on classical harmonic oscillator model, increases since the amplitude of the oscillation becomes larger.
\item The oscillation frequency is independent of temperature and approximately of chemical potential. It is intuitively comprehensible. Similar to the classical harmonic oscillator, the oscillation frequency is an intrinsic characteristic of the meson and is independent of the environmental changes. However, the larger distance between quark and antiquark, the smaller oscillation frequency.
\item Consider a meson in the plasma with non-zero temperature and chemical potential. For larger values of the chemical potential and higher temperatures, the excitation time of the meson increases. In other words, when the plasma is hotter or denser the meson falls into the final excited state more slowly.
\item Consider a plasma where its temperature and chemical potential are kept fixed. Then the more stable meson, corresponding to the smaller distance between quark and antiquark, falls into the final excited state sooner. 
\item All of the above results are confirmed for the slow and fast quench. By slow (fast) quench we mean the energy injection into the system is done slowly (rapidly).
\end{itemize}
\section{Review on the backgrounds}
The gauge-gravity duality proposes a promising approach to investigate different properties of strongly coupled field theory. Since we want to study the QGP as a strongly coupled system in the presence of non-zero chemical potential, we firstly review its corresponding holographic dual, i.e. RN-AdS background. We then extend our problem to the time-dependent case, i.e. RN-AdS-Vaidya, which is dual to the thermalization process in the strongly coupled field theory when the temperature and chemical potential simultaneously increase. 
\subsection{RN-AdS black hole background}
Here we introduce charged black hole metric which is asymptotically AdS. Consider the Einstein-Maxwell anti-de Sitter action \cite{Galante:2012pv}:
\be\label{action1}
S=-\frac{1}{{16\pi}G^{d+1}} \int d^{d+1 }x \sqrt{-g} [ {\cal {R}} - \hat{F}^2 + \frac{d (d-1)}{R^2}].
\ee%
where $G$ is Newton constant, ${\cal{R}}$ is Ricci scalar, $\hat{F}$ is field strength of the $U(1)$ gauge field and $R$ is AdS radius which we set to be one. The number of spatial directions is $d$ and it relates to the negative cosmological constant as $\Lambda=-\frac{d(d-1)}{2R^2}$. Equations of the motion obtaining from the above action are
\be\label{eq1}\begin{split}%
0&=R_{{\mu \nu}} - \frac{1}{2} g_{\mu \nu} ( {\cal {R}} -2 {\Lambda} - \hat{F}^2) - 2 \hat{F}_{\mu \lambda} {\hat{F}^{\lambda}}_{\mu}\,, \cr
0&=\frac{1}{\sqrt{-g}} \partial_{\nu} ( {\sqrt{-g}} \hat{F}^{\nu \sigma}).
\end{split}\ee%
where $\mu=0,...,d$. The solution of the equations of motion is the RN-AdS metric which can be written as \cite{Galante:2012pv}
\be\label{metric1}\begin{split} %
ds^2=& \frac{1}{z^2}[- f(z) dt^2 + \frac{1}{f(z)}dz^2 + d \vec{x}^2]\, , \cr
f(z) =& 1 - M z^d + {Q}^2 z^{2d-2},
\end{split}\ee %
and the time component of the gauge field introduced in \eqref{gauge1}. $M$ and $Q$ are the mass and charge of the RN-AdS black hole, respectively. In (\ref{metric1}), $z$ is radial coordinate and $z=0$ is the AdS boundary. In addition, $(t,\vec{x})$ are the four dimensional boundary coordinates which the field theory lives. The gauge-gravity duality indicates that the Hawking temperature of the black hole corresponds to the temperature of the QGP. For $d=4$, the temperature of the RN-AdS$_5$ black hole is
\be\label{tem}
T = \frac{1}{\pi z_h} (1 - \frac{1}{2} Q^2 z_h^6).
\ee%
Here $z_h$ is the radius of the event horizon, i.e. the smallest root of $f(z)=0$. The relation between $z_h$, $M $ and $Q$ is
\be\label{massradius}
M = \frac{1}{z_h^4} + Q^2 z_h^2.
\ee
Moreover, the time component of the gauge field is given by \cite{Galante:2012pv}
\be\label{gauge1} 
A_t=- \frac{\sqrt{3}}{2}Q z^2 + \Phi,
\ee
where $\Phi$ is a constant which plays the role of the electrostatic potential between the boundary of the bulk and the event horizon. It can be defined by applying $A_t(z_h)=0$. Thus we obtain
\be
\Phi = \frac{\sqrt{3}}{2} Q z_h^2
\ee
Gauge-gravity duality provides a correspondence between the time component of gauge field at the boundary  and chemical potential in dual boundary gauge theory \cite{Galante:2012pv, Myers:2009ij}, i.e.
\be
{\mu} =\lim_{z \rightarrow 0}A_t = \frac{\sqrt{3}}{2}Q z_h^2
\ee
in the AdS radius unit. Hence, it is easy to find that 
\be
\frac{\mu}{T} = \frac{\sqrt{3}\pi}{2} \frac{Q z_h^3}{ 1 - \frac{1}{2}Q^2 z_h^6 } \,.
\ee

\subsection{RN-AdS-Vaidya background}
Let's generalize the static background (\ref{metric1}) to the time-dependent case. It can be achieved by adding external source terms to the action (\ref{action1}). Then, the equations of motion for general $d$ are \cite{Galante:2012pv} 
\be\label{eq1}\begin{split}%
8 \pi G^{d+1} T^{(ext)}_{\mu\nu}&=R_{{\mu \nu}} - \frac{1}{2} g_{\mu \nu} ( {\cal {R}} -2 {\Lambda} - \hat{F}^2) - 2 \hat{F}_{\mu \lambda} {\hat{F}^{\lambda}}_{\nu}, \cr
8 \pi G^{d+1} J_{(ext)}^\mu&=\frac{1}{\sqrt{-g}} \partial_{\nu} ( {\sqrt{-g}} \hat{F}^{\nu \mu}).
\end{split}\ee%
The RN-AdS-Vaidya metric, in the Eddington-Finkelstein coordinates, is then given by
\be\label{metric10}\begin{split}%
ds^2= &\frac{1}{z^2}[- F({\bar{v}},z) d{\bar{v}}^2 -2 dz d{\bar{v}}+ d \vec{x}^2]\, ,
\cr
F(\bar{v},z)=&1 - M(\bar{v}) z^d + {Q(\bar{v})}^2 z^{2d-2},\cr
A_{\mu}=&- \frac{Q(\bar{v})}{c} z^{d-2} {\delta}_{\mu \bar{v}},
\end{split}\ee%
provided that
\be\label{masscharge}\begin{split}%
8 \pi G^{d+1} T^{(ext)}_{\mu\nu} &= z^{d-1}\left[ \frac{(d-1)}{2} \dot{M}(\bar{v}) - (d-1)z^{(d-2)}Q(\bar{v})\dot{Q}(\bar{v})\right] {\delta}_{\mu \bar{v}} {\delta}_{\nu \bar{v}}\,,
\cr
8 \pi G^{d+1} J_{(ext)}^\mu &= \sqrt{\frac{(d-1)(d-2)}{2}}z^{d+1}\dot{Q}(\bar{v}){\delta}^{\mu z},
\end{split}
\ee%
where $\dot{Q}(\bar{v})=dQ/d\bar{v}$ and so on. The $\bar{v}$-coordinate reduces to $t$ at the boundary, i.e. $t=\bar{v}|_{z=0}$, the time coordinate of the gauge theory. $M(\bar{v})$ and $Q(\bar{v})$ are arbitrary functions that represent how and at what rate the mass and the charge of the RN-AdS black hole increase. Various functions for $M(\bar{v})$ and $Q(\bar{v})$ have been discussed in the literature, for instance see \cite{ali-akbari:3}, and it seems that the physical results are independent of the form of the functions. Therefore, we choose 
\bea %
{\cal{I}}(\bar{v})= {\cal{I}}_f \left\{%
\begin{array}{ll}
0 & {\bar{v}}<0, \\
k^{-1}\left[{\bar{v}}-\frac{k}{2\pi}\sin(\frac{2\pi {\bar{v}}}{k})\right] & 0 \leqslant {\bar{v}} \leqslant k, \\
1 & {\bar{v}}>k ,\\
\end{array}%
\right.
\eea %
where ${\cal{I}} \in(M , Q)$. The transition time $k$ is the time interval that the mass (charge) of the black hole needs to reach its final value $M_f$ ($Q_f$). For $k\ll 1$ ($k\gg 1$) the transition time is small (large) which is usually called {\it{fast (slow) quench}}. According to gauge-gravity duality, study of the RN-AdS-Vaidya metric on the gravity side corresponds to the study of thermalization of strongly coupled QGP in the presence of the chemical potential in the gauge theory. Note that the relation between $M_f$, $Q_f$ and $z_h$ is still given by (\ref{massradius}).

\section{Expectation value of Wilson loop}
The static potential between a quark and anti-quark has been extensively studied, for example see \cite{solana, Brandhuber:1998bs}. In this section, we want to review how this static potential can be obtained from the duality point of view. To do so, we use the expectation value of the Wilson loop as a gauge invariant operator. Specifically, on the one hand, we use the time-like Wilson loop in rectangular form $\cal{C}$. One side of rectangle, $l$, is spatial corresponding to the distance between quark-antiquark pair and the other side is temporal, $\cal{T}$. If we assume that ${\cal{T}}\gg l$, meaning that the world-sheet is translationally invariant along the time direction, the expectation value of the Wilson loop is \cite{solana}
\be\label{wilson3}
\langle W({\cal{C}}) \rangle =e^{-i(2 m + V(l)){\cal{T}}},
\ee
where $m$ is the rest mass of quark (antiquark) and $V(l)$ represents static potential energy between the pair. On the other hand, the gauge-gravity duality proposes that in order to calculate the static potential in the QGP at finite chemical potential, one needs to probe the RN-AdS black hole geometry by a classical string \cite{Zeng:2013fsa}. In other words, the expectation value of the Wilson loop, in the saddle point approximation, is dual to the on-shell action of classical open string that its end points are located on the boundary with distance $l$. Therefore,
\be\label{wilson4}
\langle W({\cal{C}})\rangle =e^{i S({\cal{C}})},
\ee\label{action}
where $S({\cal{C}})$ is value of the Nambu-Goto action
\be\label{action2}
S=\frac{-1}{2 \pi \alpha'} \int d\tau d\sigma \sqrt{- \det(g_{ab})}\,.
\ee
on the rectangle ${\cal{C}}$. The tension of string is proportional to $\alpha'^{-1}\equiv {{l}_s}^{-2}$ and ${l}_s$ is the fundamental length scale of string. Also, $\tau$ and $\sigma$ parametrize the two dimensional string world-sheet. $g_{ab}=G_{\mu\nu}\frac{\partial X^\mu}{\partial \xi^a} \frac{\partial X^\nu}{\partial \xi^b}$ is the induced metric on the world-sheet. Here $G_{\mu\nu}$ and $X^\mu$ ($\xi^a=\tau,\ \sigma$) are the metric and the bulk (world-sheet) coordinates, respectively. This action, in fact, describes the dynamics of a classical string in any desired geometry. As we will see, using (\ref{wilson3}), (\ref{wilson4}) and (\ref{action2}), the static potential can be found. Note that, the rest mass $m$ is equal to $\frac{\sqrt{\lambda}}{2\pi}\int_{\epsilon}^{z_h}\,\frac{dz}{z^2}$ \cite{solana}, where $\epsilon$ is IR regulator in the gravity theory and according to UV/IR connection, it corresponds to the UV cut-off in the gauge theory. 

In order to calculate the $S({\cal{C}})$ in the RN-AdS black hole background (\ref{metric1}), we parametrize the two-dimensional world-sheet of the string as $\tau= t$, $\sigma=x_3\equiv x$. All bulk coordinates, except $z$ and $x$, are chosen to be constant and therefore the shape of the string is described by $z=z(x)$. Hence the action (\ref{action2}) reduces to
\be\label{staticaction}
S=-\frac{{\cal{T}}}{2\pi\alpha'}\int_{-\frac{l}{2}}^{\frac{l}{2}}\, dx\frac{1}{z^2}
\sqrt{{z^\prime}^2+{f}(z)},
\ee
where $z'\!=\!dz/dx$. Since the Lagrangian does not depend explicitly on $x$, the associated Hamiltonian is a constant of motion. After some simple algebra, we get
\be\label{staticz}
z'(x)=\pm\frac{{z_\ast}^{2}~f(z)}{\sqrt{{f}(z_\ast)}~z^{2}}~
\sqrt{1-\frac{{f}(z_\ast)}{{f}(z)}\left(\frac{z}{{z_\ast}}\right)^{4}},
\ee
where $z=z_\ast$ at $z'(x)=0$. Using the new coordinate $y=z_*/z$ and the explicit form of $f(z)$, it turns out
\begin{align}\label{staticx}
\int_{\pm\frac{l}{2}}^x\, dx&=\mp z_{\ast} \sqrt{1-y_h^{4}+{Q_f}^2 {z_\ast}^6(1-{y_h}^{-2})} \nonumber \\ & \times
\int_1^y\frac{dy}{\sqrt{\left[y^{4}-1+{Q_f}^2 {z_\ast}^6({y}^{-2}-1)\right]\left[y^{4}-y_h^{4}+{Q_f}^2 {z_\ast}^6({y}^{-2}-{y_h}^{-2})\right]}},
\end{align}
where $y_h=z_*/z_h$. The on-shell action can be then found by inserting (\ref{staticz}) and \eqref{staticx} into (\ref{staticaction}). Furthermore, by applying (\ref{wilson3}) and (\ref{wilson4}) the potential energy between quark and anti-quark will be obtained. Notice that, in order to get a finite value for the potential energy, we need to subtract the rest mass of the quarks from the on-shell action and we therefore have
\begin{align}\label{finallstatic}
V = \frac{1}{\pi\alpha^\prime} \bigg[\frac{1}{z_\ast}& \int_1^\infty\,dy 
\left(\sqrt{\frac{y^{4}-y_h^{4}+{Q_f}^2 {z_\ast}^6({y}^{-2}-{y_h}^{-2})}{y^{4}-1+{Q_f}^2 {z_\ast}^6({y}^{-2}-1)}}-1\right) 
-\left(\frac{1}{z_{\ast}}-\frac{1}{z_h}\right)\bigg].
\end{align}
Before closing this part, two points should be noted. First, as we will see, the equation \eqref{staticx} will be utilized as an initial condition for the time evolution of classical string in the RN-AdS-Vaidya background.
Second, in the next section, we observe that the expectation value of the time-dependent Wilson loop oscillates around the static potential resulted in (\ref{finallstatic}). 

Now our goal is to calculate the evolution of expectation value of Wilson loop in the RN-AdS-Vaidya geometry. In this geometry, the two-dimensional world-sheet of the string is no longer translationally invariant in the time direction. Thus the condition ${\cal{T}}\gg l$ does not valid and therefore in this case, the expectation value of the Wilson loop (\ref{wilson3}) can be written as
\be\label{timewilson}
\langle W({\cal{C}})\rangle =e^{-i \int dt\ {\cal{W}}(t)}\,,
\ee%
where, using gauge-gravity duality, ${\cal{W}}(t)$ is the on-shell action of string without integrating over $t$-coordinate. It is important to note that although ${\cal{W}}$ is a function of different parameters in the gauge theory, such as temperature, chemical potential and distance $l$, we only show its time dependence, explicitly. As before, $ {\cal{W}}(t)$, or equivalently on-shell string action, diverges and in order to regularize this divergence, similar to the static case, we have
\be\begin{split}\label{regul}%
{\cal{W}}_R(t) = {\cal{W}}(t) - 2 m \equiv \int d\sigma \left(\sqrt{- \det (g_{ab})}\right)_{\rm{on-shell}} - 2 m.
\end{split}\ee%
The subscript $"R"$ in (\ref{regul}) refers to the regularized version of ${\cal{W}}(t)$. To calculate the string on-shell action, similar to \cite{ishii}, we choose the null coordinates $(u,v)$ to parametrize the two-dimensional world-sheet of the string and therefore all the coordinates on the world-sheet depend on $u$ and $v$. Then we choose the following ansatz:
\be\label{ansatz7}%
\bar{v}=V(u,v)~ , ~ z=Z(u,v) ~,~ x_3=X(u,v)\,.
\ee%
Note that due to rotational symmetry in spatial coordinates, there is no difference among them. Substituting this ansatz into the Nambo-Goto action (\ref{action2}), the equations of motion can be found. Hence, we finally have
\be\label{eom}\begin{split}%
V_{,uv}&=
\left( \frac{F_{,Z} }{2} -\frac{F}{Z}\right) V_{,u} V_{,v} + \frac{1}{Z} X_{,u} X_{,v}\, ,
\cr
Z_{,uv}&=\left(\frac{F^2}{Z} - \frac{F }{2} F_{,Z} -\frac{1}{2} F_{,V}\right) V_{,u} V_{,v} 
+ \left(\frac{F}{Z} -\frac{F_{,Z}}{2}\right) \left(Z_{,u} V_{,v}+Z_{,v} V_{,u}\right)+ \frac{2}{Z} Z_{,u} Z_{,v} - \frac{F}{Z} X_{,u} X_{,v}\, ,
\cr
X_{,uv}& = \frac{Z_{,u} X_{,v} + Z_{,v} X_{,u}}{ Z}.
\
\end{split}\ee%
Since $u$ and $v$ are null coordinates, we need to impose two constraints corresponding to $g_{uu}=0$ and $g_{vv}=0$. Thus, two constraint equations turn out to be
\be\label{cons}\begin{split}%
C_1 = \frac{1}{Z^2} (F V_{,u}^2 + 2 V_{,u} Z_{,u} - X_{,u}^2) & = 0,\cr
C_2 = \frac{1}{Z^2} (F V_{,v}^2 + 2 V_{,v} Z_{,v} - X_{,v}^2 ) & = 0\,.
\end{split}\ee%
In order to solve the equations of motion (\ref{eom}) subject to constraint equations (\ref{cons}), the suitable boundary and initial conditions are essential. In the appendices \ref{a1} and \ref{b1}, we obtain the appropriate boundary and initial conditions, respectively. After solving the equations of motion \eqref{eom}, one can easily find the time evolution of ${\cal{W}}_R(t)$ when the other parameters are kept fixed.
\begin{figure}[ht]
\begin{center}
\includegraphics[width=66 mm]{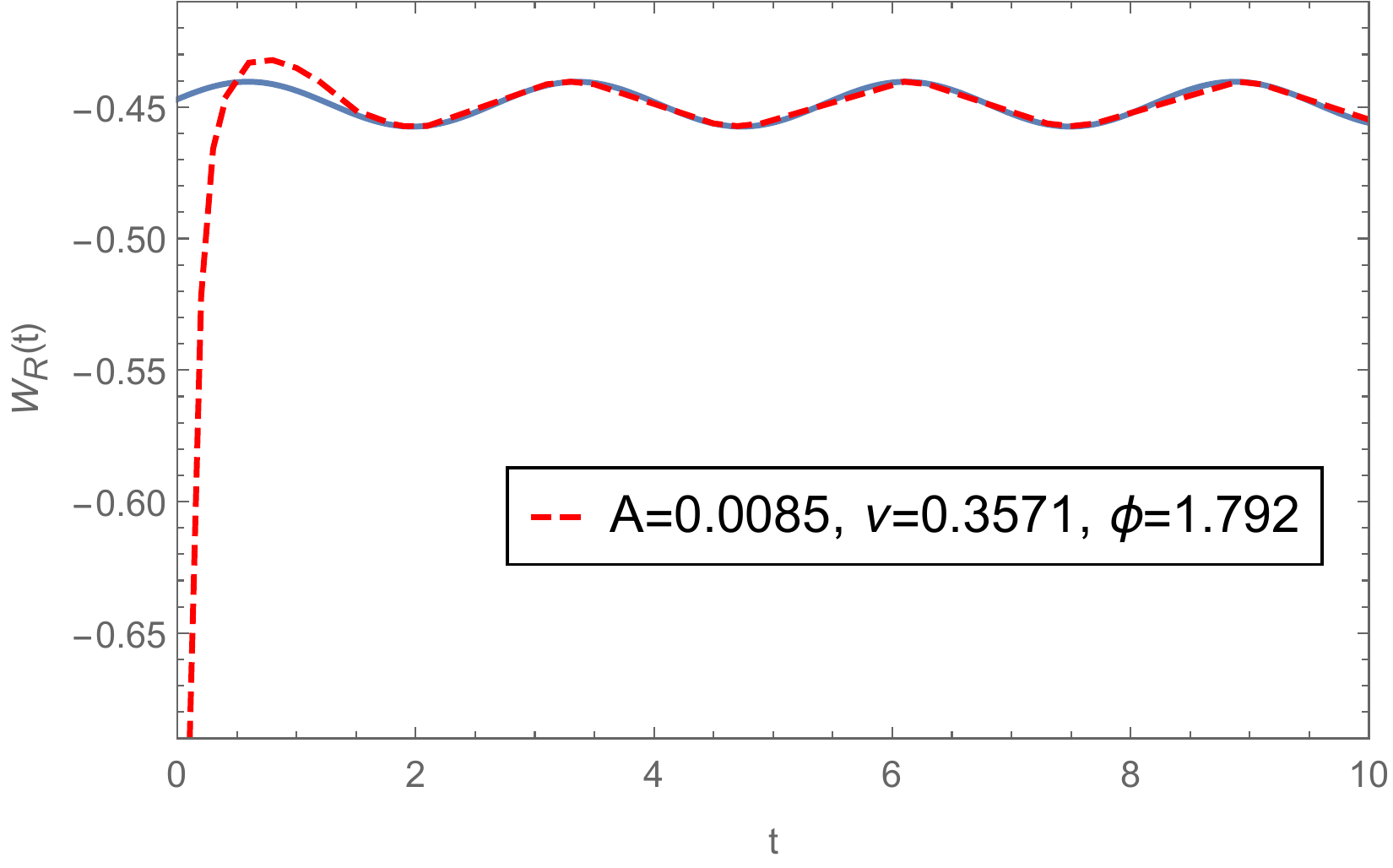}
\hspace{2 mm}
\includegraphics[width=66 mm]{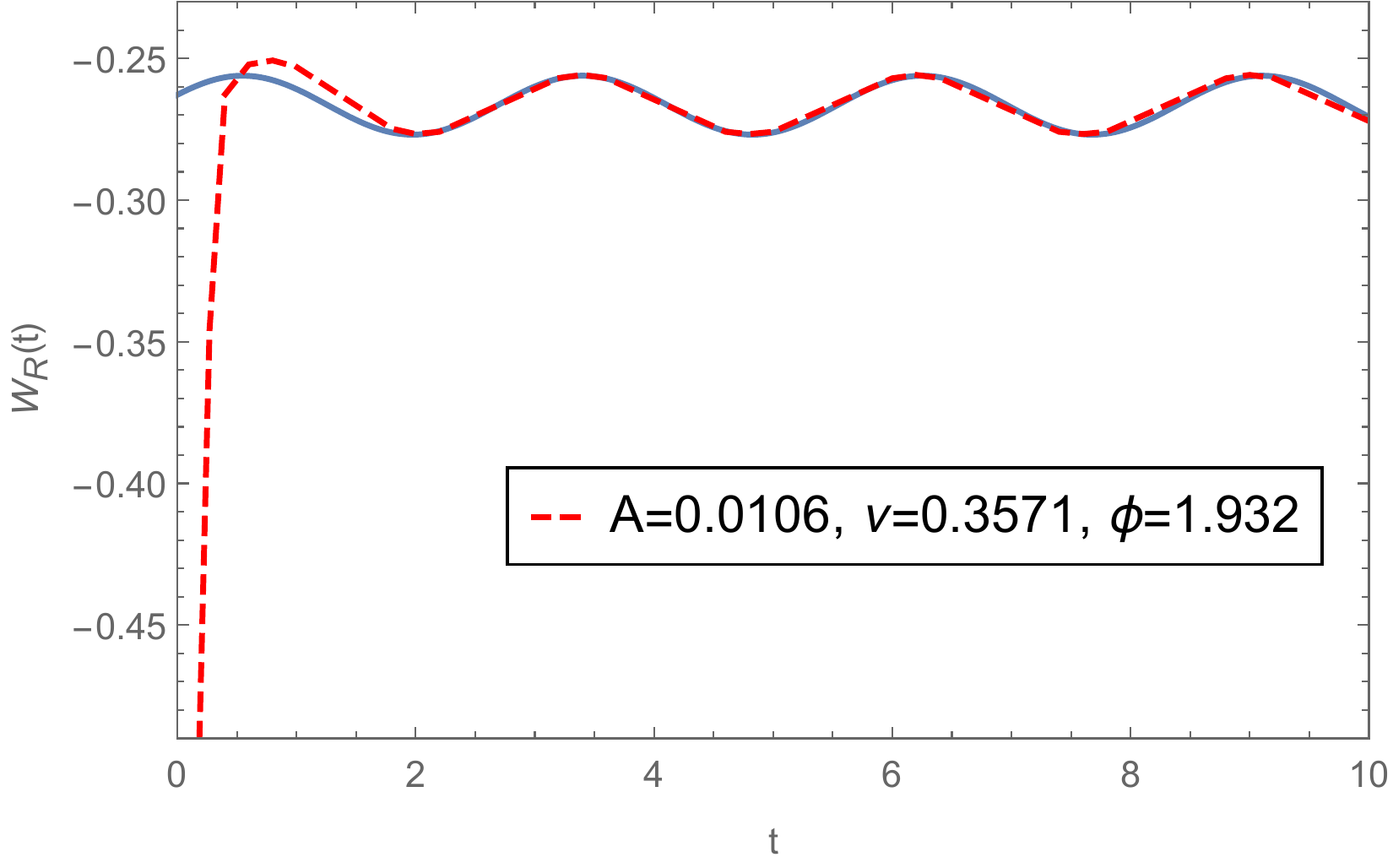}
\caption{ ${\cal{W}}_R(t)$ in terms of the boundary time $t$ for $T_f=0.1593$, $l=1$ and $k=0.3$. The final chemical potential for the left (right) panel is $\mu_f=0$ ($0.3076$). 
The static potentials for left (right) graph is $V=-0.4488$ ($-0.2664$). The dashed red curve shows our numerical results and the blue sine curve is the fitted function \eqref{fit2}. 
\label{f1}}
\end{center}
\end{figure}%
\begin{figure}[ht]
\begin{center}
\includegraphics[width=66 mm]{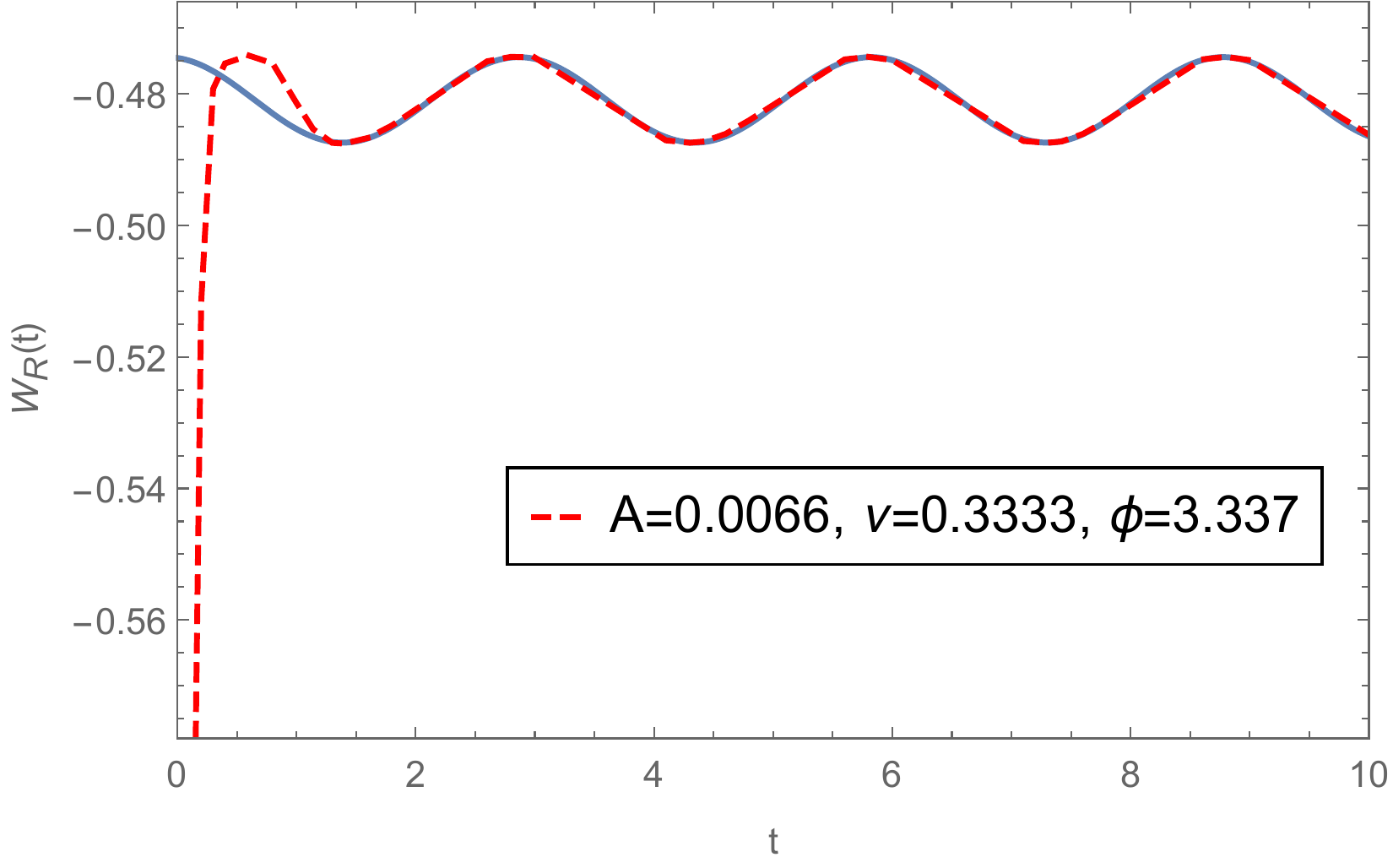}
\hspace{2 mm}
\includegraphics[width=66 mm]{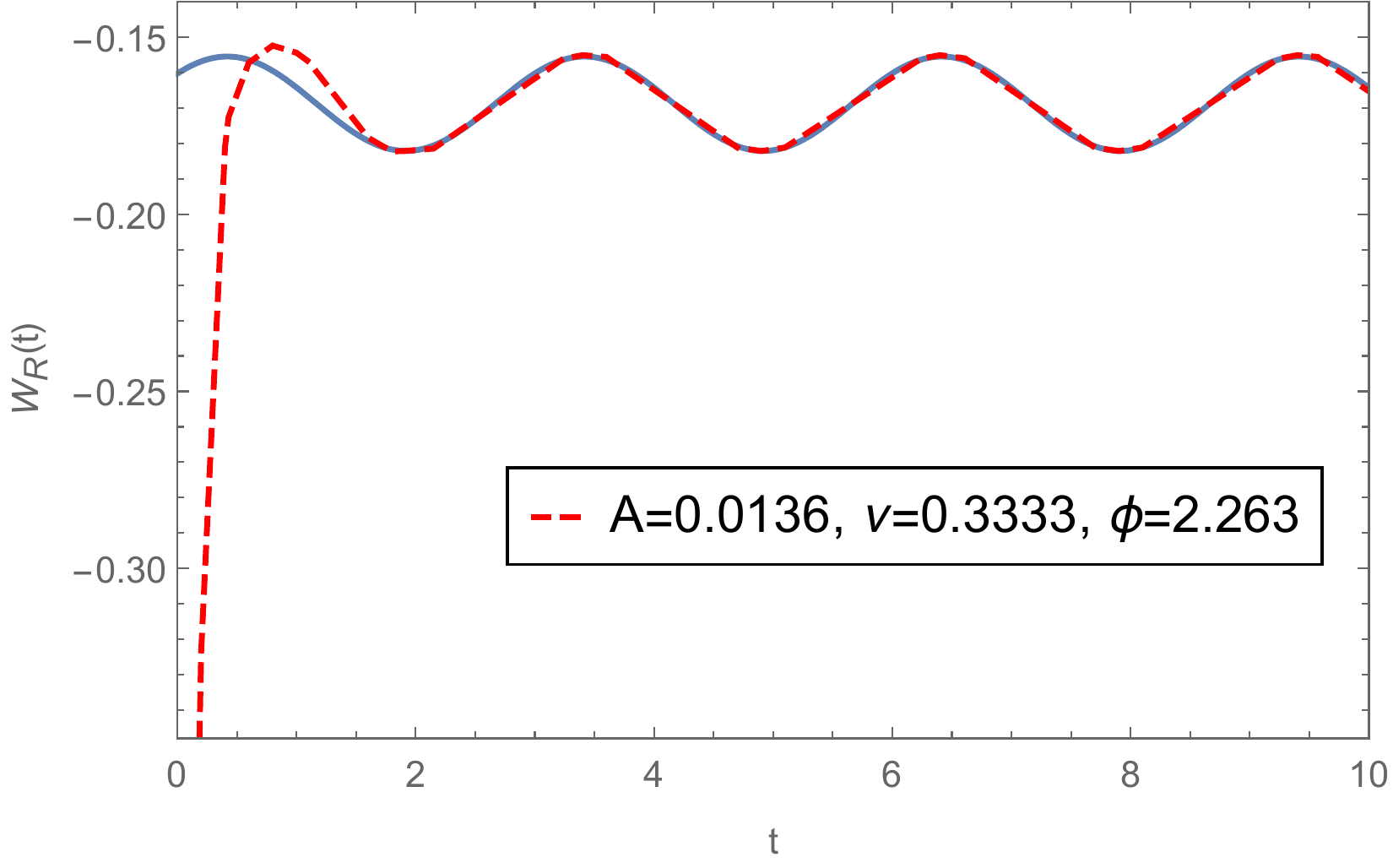}
\caption{${\cal{W}}_R(t)$ in terms of boundary time $t$ for $\mu_f=0.4503$, $l=1$ and $k=0.3$. The final temperature for the left (right) panel is $0.0731$ ($0.1428$). The static potential is $V(l)=-0.4809 (-0.1686)$ for left (right) graph. The dashed red curve shows our numerical results and the blue sine curve is the fitted function \eqref{fit2}.
\label{f3}}
\end{center}
\end{figure}%
\begin{figure}[ht]
\begin{center}
\includegraphics[width=67 mm]{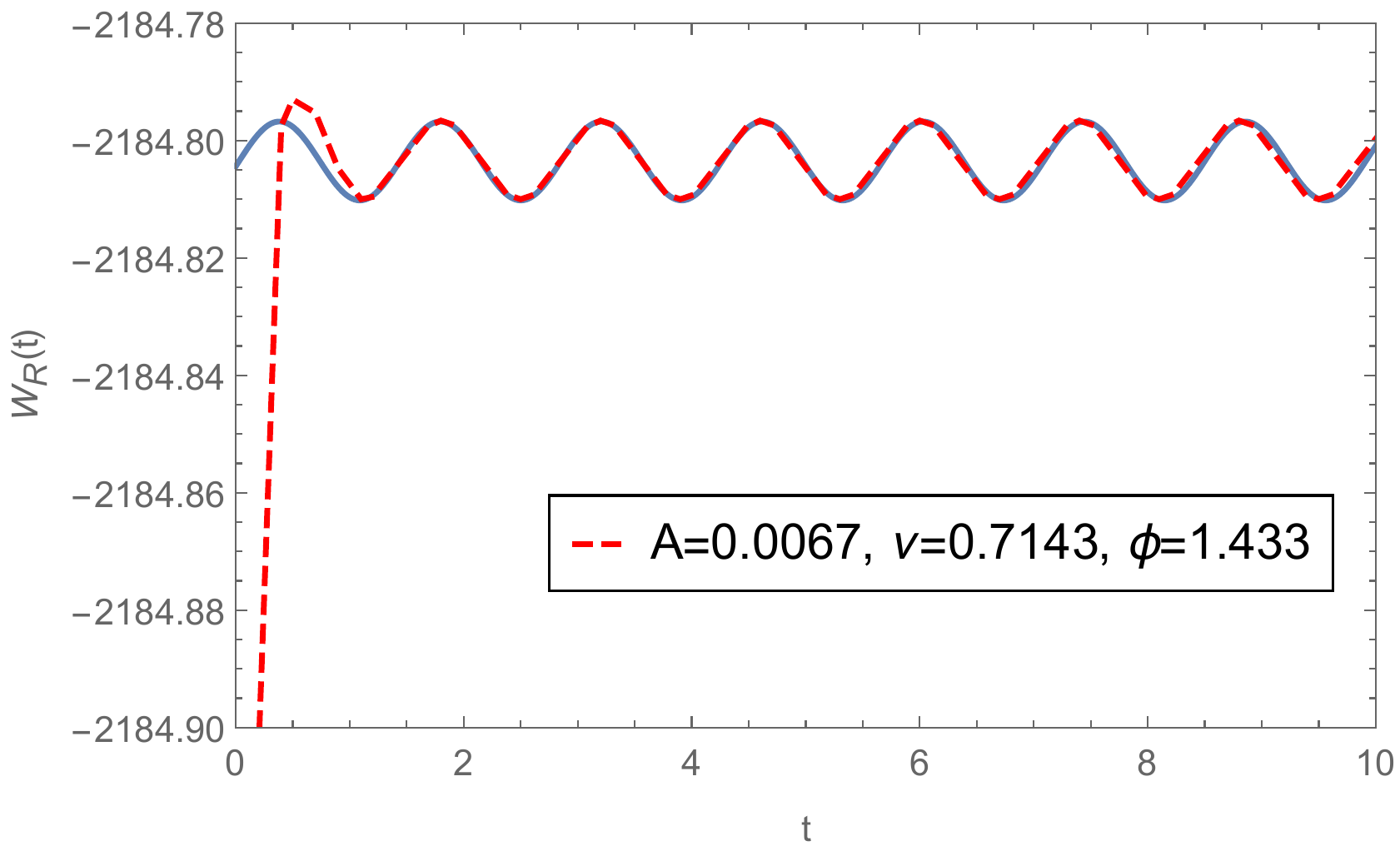}
\hspace{2 mm}
\includegraphics[width=63 mm]{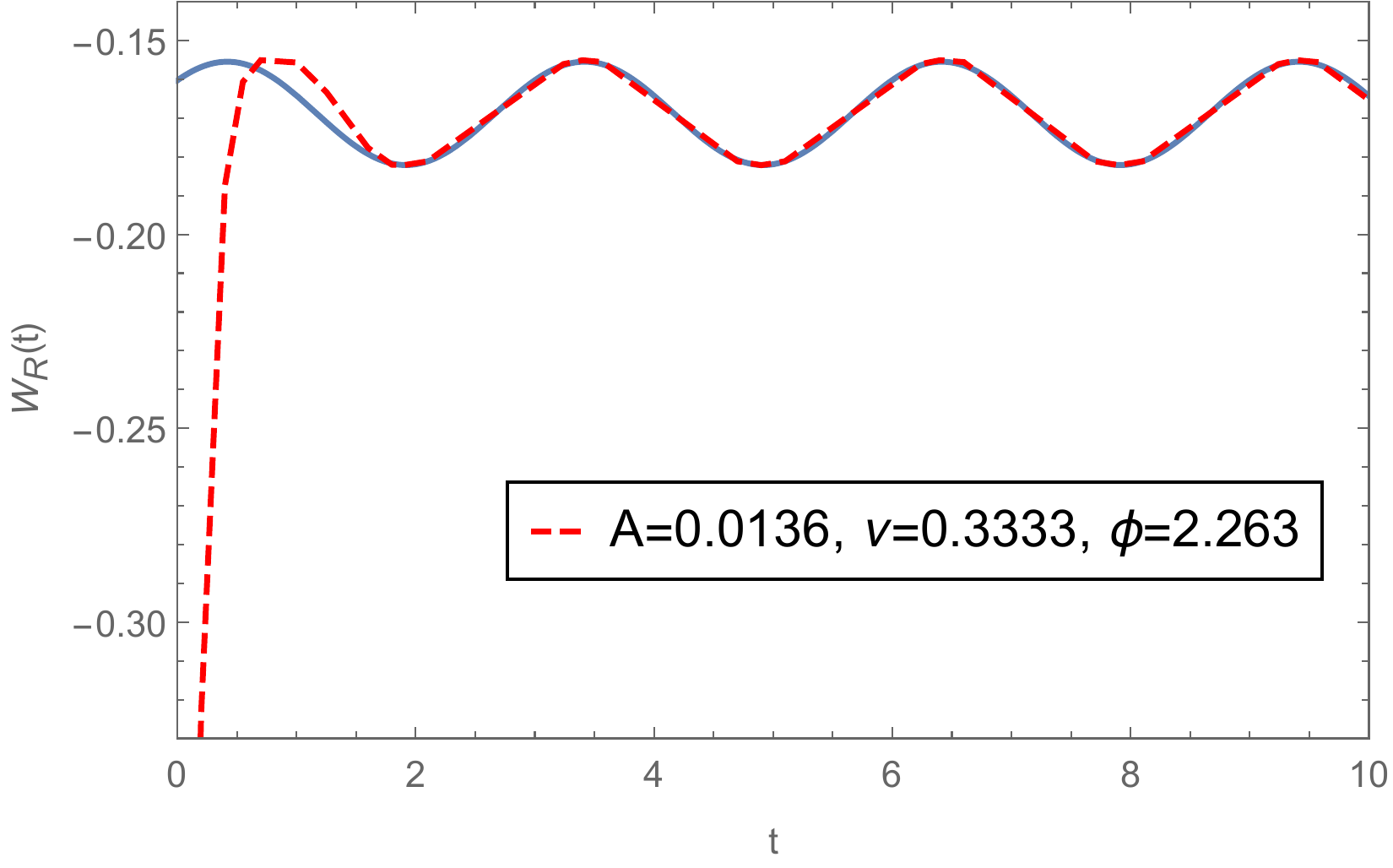}
\caption{ ${\cal{W}}_R(t)$ in terms of the boundary time, for $T_f=0.1428$, $\mu_f=0.4503$ and $k=0.3$. The distance between the pair is $l=0.5$ ($l=1$) for the left (right) panel. 
The static potential is $V(l)=-2184.8 (-0.1686$) for left (right) panel, respectively. The dashed red curve shows our numerical results and the blue sine curve is the fitted function \eqref{fit2}.
\label{f4}}
\end{center}
\end{figure}%

\section{Numerical results}
As we have already mentioned and it is approved by the figures \ref{f1}, \ref{f3} and \ref{f4}, the expectation value of the Wilson loop oscillates around the static potential. In fact, before injecting energy ($t<0$), the quark-antiquark pair is in ground state. However, as the energy injection is started, the temperature and chemical potential increases and therefore the pair is excited. When the energy injection ceases, the pair relaxes toward a final excited state with specify frequency and amplitude of oscillation. In other words, these oscillations can be interpreted as if the energy injection puts the meson in the excited state \cite{ishii}. We are now interested in investigating the effect of three parameters final chemical potential $\mu_f$, final temperature $T_f$ and the distance between the quark and antiquark $l$ on the characteristic of oscillation, i.e. frequency and amplitude. Thus our results are classified into three categories:
\begin{itemize}
\item {\textbf{Fixed temperature and distance:}} The effect of chemical potential on the expectation value of the Wilson loop is easily seen in the figure \ref{f1}. In the left (right) panel the final chemical potential is equal to $0\ (0.3076)$. An enhancement in the final chemical potential increases the amplitude of the oscillation.  However the frequency of the osillation is almost constant. Therefore, comparing to simple harmonic oscillator, in this case, in the QGP with non-zero matter the energy of the excited pair becomes larger. Moreover, the value of the static potential indicates that the pair is less stable in the presence of chemical potential.
\item {\textbf{Fixed distance and chemical potential:}} By raising the final value of the temperature of the QGP, the amplitude of the oscillation increases, see figure \ref{f3}. However, the value of the frequency is independent of the final temperature in agreement with the numerical results of \cite{Hajilou:2017sxf}. In this case, at higher temperatures the energy of the excited pair increases and quark-antiquark bound state is less stable.
\item {\textbf{Fixed chemical potential and temperature:}} In figure \ref{f4}, with increasing the distance between a quark and antiquark, the amplitude (frequency) of oscillation increases (decreases). Although the energy of the excited pair becomes greater for larger values of $l$, this bound state is less stable.
\end{itemize}
The effect of the chemical potential on the equilibration (or thermalization) time has been extensively studied in the literature, for instance see \cite{Camilo:2014npa, Zeng:2013fsa, Galante:2012pv}. It is shown that larger values of the chemical potential yield larger equilibration times. However, in \cite{Caceres:2012em, Zeng:2013fsa}, the results indicate that the equilibration time becomes larger or shorter depending on the value of $\mu/T$. More recently, equilibration of a dynamical scalar operator is considered in the charged QGP during its equilibration \cite{Ebrahim:2017gvk} and the numerical outcomes show that the equilibration time can be a decreasing or increasing function of the chemical potential. Therefore, the conclusion may be as follows: although the excitation time increases for larger values of the chemical potential in our case, it should be instructive to check our outcomes in other gauge theories with holographic dual to find a general behavior.

Up to now, we investigate the response of the system to the time-dependent change in temperature and chemical potential. In fact, the response of the system is described by the behavior of the expectation value of the Wilson loop, ${\cal{W}}_R(t)$, in terms of boundary time. We observe that the expectation value oscillates around the static potential with a specific value of frequency, $\nu$, and amplitude of oscillation, $A$. The characteristics of oscillation depend on the final values of the chemical potential and temperature as well as the distance between the pair. Now, the question we would like to answer is, at which time the expectation value of the Wilson loop starts oscillating around the static potential? And we call it {\it{excitation time}}. To do so, we define the following function:
\be\label{fit2}
{\cal{W}}_f(t) = A\cos (2 \pi \nu t +{\phi})\, ,
\ee
where $A$, $\nu$ and $\phi$ can be found from ${\cal{W}}_R(t)$ at asymtotic times. We then define a time-dependent function
\bse\begin{align}
\label{991} {\epsilon}_{\mu}(t) = \mid\frac{{\cal{W}}_R(t) - {\cal{W}}_f(t)}{{\cal{W}}_R(t)}\mid_{T,l}, \\
\label{992} {\epsilon}_{T}(t) = \mid\frac{{\cal{W}}_R(t) - {\cal{W}}_f(t)}{{\cal{W}}_R(t)}\mid_{\mu,l}, \\
\label{993} {\epsilon}_{l}(t) = \mid\frac{{\cal{W}}_R(t) - {\cal{W}}_f(t)}{{\cal{W}}_R(t)}\mid_{\mu,T},
\end{align}\ese 
where in \eqref{991} $T$ and $l$ are kept fixed while $\mu$ changes and so on. Thus the excitation time, $t_{ex}$, is defined as the time which satisfies $\epsilon(t_{ex})<5\times 10^{-6}$ and $\epsilon(t)$ stays below this limit afterwards.

\begin{figure}[ht]
\begin{center}
\includegraphics[width=66 mm]{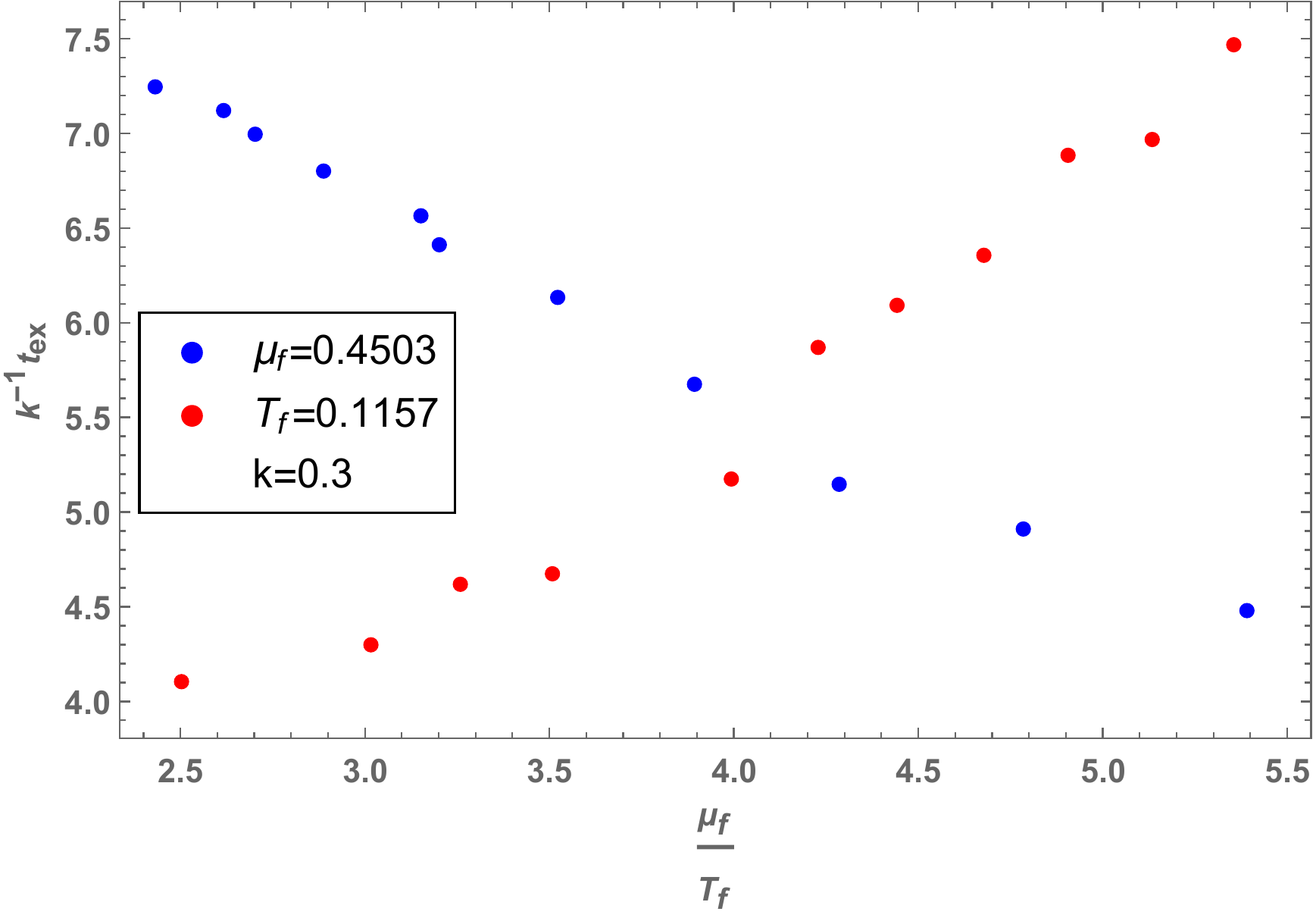}
\hspace{2 mm}
\includegraphics[width=66 mm]{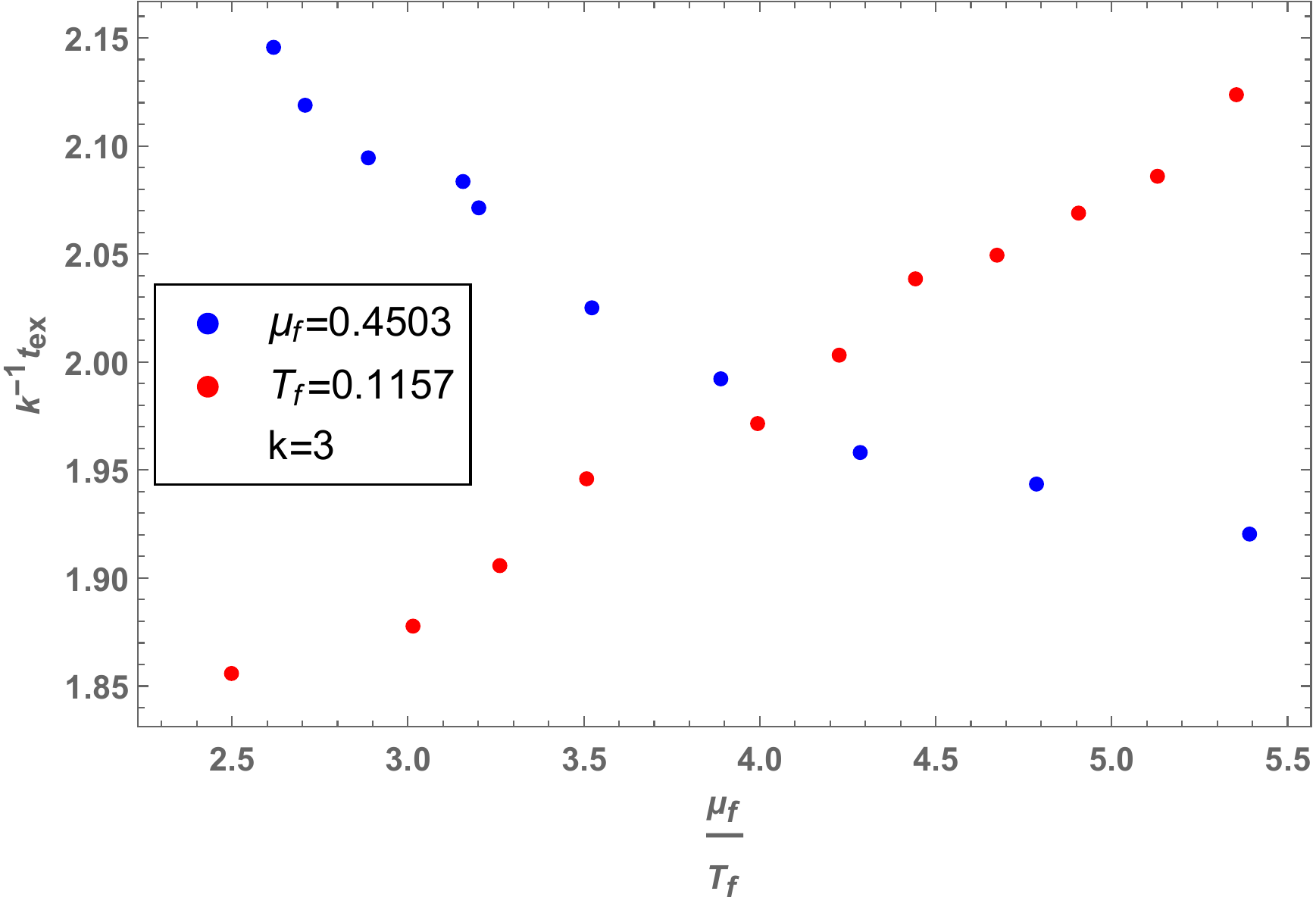}
\caption{The excitation time in terms of $\mu_f/T_f$ for $l=1$. 
\label{figure36}}
\end{center}
\end{figure}%

In figure \ref {figure36}(left), blue points, the final temperature of the system decreases while the other parameters contributing to the time evolution of the expectation value are kept fixed. As it is clearly seen, at higher temperatures the meson falls into the excited state later. In other words, the higher temperature, the larger (rescaled) excitation time, $(k^{-1})t_{ex}$. This result can be intuitively understood: at higher temperatures the thermal fluctuation becomes more significant and thus its effect prevents the meson from sooner relaxation.
In the same figure, red points, since we would like to investigate the effect of the chemical potential on the excitation time, the chemical potential is varied and the other parameters are kept fixed. Similar to the previous case, this figure indicates that the rescaled excitation time and chemical potential increase with respect to each other. As a matter of fact, for larger values of the temperature and chemical potential, the rescaled excitation time becomes larger. In short, the effect of temperature on the rescaled excitation time is similar to the chemical potential. 
Apart from the fast quench, in the right panel, the rescaled excitation time is plotted as a function of $\mu_f/T_f$ for the case of slow quench, i.e. $k=3$. It is obviously seen that in both cases the behavior of excitation times is the same. However, $k^{-1} t_{ex}$ is larger for the case of fast quench.  
\begin{figure}
\includegraphics[width=66 mm]{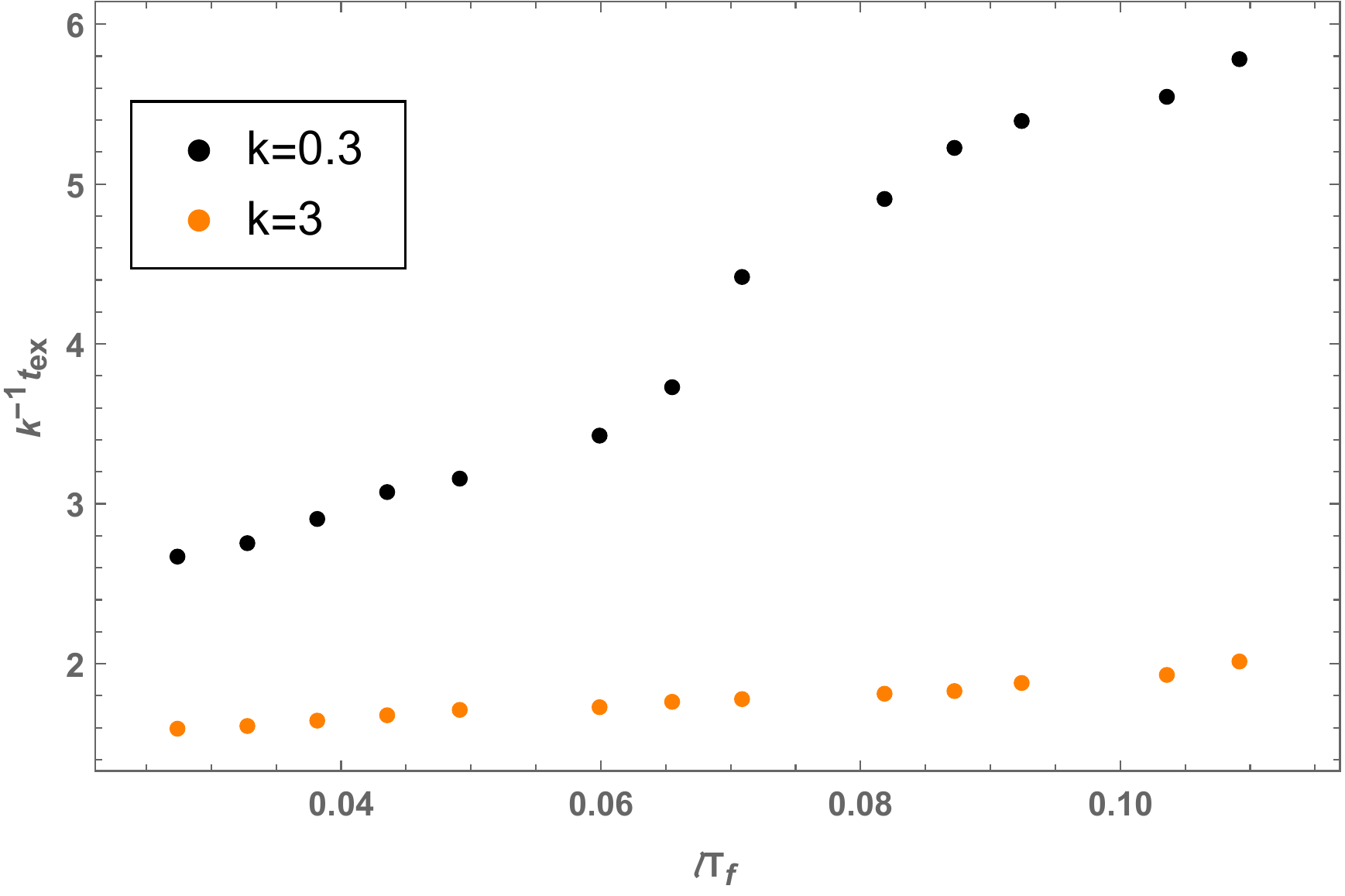}
\caption{The excitation time in terms of $lT_f$ for $T_f=0.1091$ and $\mu_f=0.4811$. 
\label{figure37}}
\end{figure}%

In figure \ref{figure37}, the distance $l$ is increased while the other parameters are kept fixed. In contrast to the chemical potential and temperature, the distance is an intrinsic characteristic of the meson in the plasma. This figure shows that for larger value of distance $l$, meaning that the meson is less stable, the excitation time increases. Although for both quenches the rescaled excitation time behaves similarly, the value of $k^{-1}t_{ex}$ is larger for the fast quench. In other words, this figure indicates that the more stable quark-antiquark bound states are excited sooner. It may be related to the screening of the force between color charges of the quark and antiquark due to the presence of the medium since they, quark and antiquark, can not communicate, easily. Notice that the distance between quark and antiquark can not be too large since the bound state becomes unstable.
\appendix
\section{Boundary condition}\label{a1}
Here, we state the boundary conditions for solving the equations of motion (\ref{eom}). By applying the diffeomorphism invariance on the two-dimensional world-sheet of the string, one of the endpoints of string  on the boundary can be fixed to be at $u=v$ and the other one at $u=v+L$. Thus, for the coordinates $Z$ and $X$, the boundary conditions are
\be\begin{split}%
Z|_{u=v}&=0~;~~~~X|_{u=v}=\frac{-l}{2},\cr
Z|_{u=v+L}&=0~;~~~~X|_{u=v+L}=\frac{l}{2}.
\end{split}\ee
The other boundary conditions can be obtained by expanding $V(u,v)$, $Z(u,v)$ and $X(u,v)$ about the point $u=v$ at the boundary. Therefore, one gets
\bea%
V(u,v)&=&V_0(v) + V_1(v) (u-v) + ...~,~~~~~~~~\\ 
Z(u,v)&=&Z_1(v) (u-v) +Z_2(v) (u-v)^2+ ...~,~~~~~~~~\\
X(u,v)&=&\frac{-l}{2} + X_1(v) (u-v) + ...~.~~~~~~~~
\eea%
The same expansion is done for the other point $u=v+L$ at the boundary. By inserting the above equations into the evolution equations (\ref{eom}) and demanding the regularity condition at $u=v$ and $u=v+L$, one can obtain the remaining boundary conditions. Notice that the results should be consistent with the constraint equations (\ref{cons}). Finally, at the boundary, i.e. $u=v$, conditions are:
\bse\begin{align}%
\label{bc1}
V(u,v)&=V_0(v) + {\cal{O}}\left((u-v)^5\right),\\ 
\label{bc2} 
Z(u,v)&=\frac{\dot{V}_0(v)}{2} (u-v) + \frac{\ddot{V}_0(v)}{4} (u-v)^2+\frac{\dddot{V}_0(v)}{12} (u-v)^3 + {\cal{O}}\left((u-v)^4\right),\\ 
X(u,v)&=\frac{-l}{2} + {\cal{O}}\left((u-v)^3\right), 
\end{align}\ese%
which indicate that
\be%
Z_{,uv}|_{u=v} = 0,\ 2 Z_{,u}|_{u=v} = \dot{V}_0(v),
\ee%
where $\dot{V}(v)=\frac{dV(v)}{dv}$. The same results are obtained for $u=v+L$. We refer the interested reader to \cite{ishii} for more details. 

\section{Initial condition}\label{b1}
Using the equation (\ref{staticz}) and constraint equations (\ref{cons}), we get the initial condition for $Z$, $V$ and $X$. We set $f(z)=1$ and replace $x$ and $z$ with the capital forms. Considering $V_{,v}>0$ at the boundary, (\ref{bc1}) and (\ref{bc2}), it is easy to find that $Z_{,u}>0$ and $Z_{,v}<0$. By imposing $X_{,u}|_{Z=0}=X_{,v}|_{Z=0}=0$ and the above conditions on the constraint equations (\ref{cons}), we get
\bea%
\label{eqV1}
V_{,u}
&= Z_{,u} \bigg(-1+\sqrt{1+\big(\frac{dX}{dZ}\big)^2}\bigg) ,~\\
\label{eqV2}
V_{,v}
&= Z_{,v} \bigg(-1-\sqrt{1+\big(\frac{dX}{dZ}\big)^2}\bigg).
\eea%
Now by taking the derivative of (\ref{eqV1}) and(\ref{eqV2}) with respect to $v$ and $u$ respectively and noting that $V_{,uv}=V_{,vu}$, it turns out
\be\label{equ}%
\bigg( Z_{,u} \sqrt{1+\big(\frac{dX}{dZ}\big)^2} \bigg)_{,v}=0.
\ee%
Replacing $\frac{dX}{dZ}$, from (\ref{staticz}), into the above equation we get the initial condition for $Z(u,v)$
\be%
\label{solZ}
Z \, _2F_1\bigg(\frac{1}{2},\frac{1}{4};\frac{5}{4};\frac{Z^{4}}{{Z_*}^{4}}\bigg) = \phi(u) - \phi(v),
\ee%
where $\phi(y)$ is an arbitrary function and we choose $\phi(y)=y$ \cite{ishii}. 

The initial condition for $X(u,v)$ is obtained by integrating (\ref{staticz}). We then have
\be%
X(u,v) = \frac{l}{2} - \frac{Z^3}{3 {Z_*}^{2}} \, _2F_1\left(\frac{1}{2},\frac{3}{4};\frac{7}{4};\frac{Z^{4}}{Z_*^{4}}\right).
\ee%
Finally, the initial condition for $V(u,v)$ can be obtained from (\ref{eqV1}) and (\ref{eqV2})
\bea%
V(u,v)&=- Z \left(1-\, _2F_1\left(\frac{1}{2},\frac{1}{4};\frac{5}{4};\frac{Z^{4}}{Z_*^{4}}\right)\right)+\chi(v),~~~~~~~\\
V(u,v)&=- Z \left(1+\, _2F_1\left(\frac{1}{2},\frac{1}{4};\frac{5}{4};\frac{Z^{4}}{Z_*^{4}}\right)\right)+{\tilde{\chi}}(u),~~~~~~~
\eea%
where $\chi$ and $\tilde{\chi}$ are some kind of arbitrary functions. Also, Using (\ref{solZ}) and equalize the above equations, we can get
\be%
\chi(v) = 2 \phi(v),~~~~~{\tilde{\chi}}(u) = 2 \phi(u).
\ee%
One can find more details in \cite{ishii}.
%

\end{document}